\documentstyle[epsf,prd,aps]{revtex}
\begin{document}
\draft
\title
{The Huygens principle and cosmological gravitational waves in the Regge-Wheeler gauge.}
\author{ Edward Malec and  Grzegorz Wyl\c e\.zek }
\address{ Institute of Physics,  Jagiellonian University,
30-059  Cracow, Reymonta 4, Poland.}

\maketitle
\begin{abstract}
We study the propagation of axial gravitational waves in Friedman universes.
The evolution equation is obtained in the Regge-Wheeler gauge. The gravitational waves
obey the Huygens principle in the radiation dominated era, but in the matter dominated
universe their propagation depends on their wavelengths, with the scale fixed  essentially
by the Hubble radius. Short waves   practically satisfy the Huygens principle while long
waves can  backscatter off the curvature of a spacetime.
\end{abstract}

\section{Introduction.}

Information about the topological type of the Universe can be
obtained through the analysis of the Hubble relation \cite{Weinberg}.
One of the key  and well known facts underlying   derivation
of the Hubble law is that electromagnetic waves
propagate along null cones, that is that they satisfy the Huygens principle
in Friedman universes \cite{Hadamard}. It is    known that gravitational
waves do not obey the Huygens principle -- they can backscatter on the
curvature of the spacetime and a part of the radiation would come  with a delay.
The observation of the effects of the backscatter  opens, in principle at least,
a new local way to see the global topology in a way analogous to effects known in
the Schwarzschild spacetime  (\cite{back} -- \cite{kark}) .

The main goal of this paper is to explain whether and when these effects
can manifest. We consider gravitational waves propagating in a fixed
Friedman universe. The gravitational waves correspond to the so-called
axial modes and they fulfill the Regge-Wheeler gauge condition \cite{RW}. The propagation equations
become particularly simple in this gauge -- they reduce to a single equation.
 This analysis with the  Regge-Wheeler condition  is novel  in
the cosmological literature, up to our knowledge. The form of the equation immediately
implies that axial modes do satisfy the Huygens principle in the radiation epoch.
On the other hand, the principle  might well be broken  in the matter-dominated epoch,
depending on the  wavelength   of the gravitational radiation. The  wavelength
scale $\lambda $ is given by the average energy density $\rho $,
$\lambda  \propto 1/\sqrt{\rho }$;
under suitable conditions   $\lambda $  is of the order
of the  Hubble radius
-- that is, of the radius of the visible part of the universe.  The effects of backscattering
are negligible in the short-range spectrum with wavelengths much smaller
than $\lambda  $ and they can possibly
manifest for longer waves.
That would mean that eventual effects cannot be observed by
detectors of gravitational waves  but they might be  responsible for
some superhorizon perturbations.

\section{Main equations.}

Let $ds^2=a^2\left( -d\eta^2+dr^2+\sigma^2d\Omega^2\right)  $
be the line element
of a Friedman cosmological model. Here $\eta $ is the conformal time
and the coordinate  radius $\sigma =r, \sin r$ and
$\sinh r$ for the flat ($k=0$),
closed ($k=1$ and open ($k=-1$) universes, respectively.
The axial perturbations of this metric in the Regge-Wheeler gauge
\cite{RW} are
defined as follows
\begin{eqnarray}
g_{0\phi }&=&h_0(\eta ,r) \sin \theta \partial_{\theta } Y_{l0}
\nonumber \\
g_{1\phi }&=&h_1(\eta ,r) \sin \theta \partial_{\theta } Y_{l0};
\label{1.1}
\end{eqnarray}
strictly saying, this a $\phi $-independent subset of a larger class,
but the general axial modes conform to the same dynamical equations
as the ones  displayed below.
The linearized Einstein equations (for $l\ge 2$, applying arguments
of  \cite{RW} and \cite{Zerilli}) are
\begin{eqnarray}
\partial_{\eta }h_0&=& \partial_r h_1 ,
\label{1.2a}
\end{eqnarray}
\begin{eqnarray}
&&\partial^2_{\eta }h_1-\partial_{\eta }\partial_rh_0+2\partial_{\eta }h_0
{\partial_r\sigma\over \sigma }-2\partial_{\eta }
\left( h_1{\partial_{\eta } a\over a}\right) + \left[ l\left( l+1\right)
-2\right] {h_1\over \sigma^2}=0,
\label{1.2b}
\end{eqnarray}
\begin{eqnarray}
&&\partial^2_rh_0-\partial_{\eta }\partial_rh_1-2\partial_{\eta }h_1
{\partial_r\sigma\over \sigma }+ 2 {\partial_{\eta } a\over a\sigma^2}
\partial_r\left( \sigma^2h_1\right) -  l\left( l+1\right)
{h_0\over \sigma^2}+4kh_0=0.
\label{1.2c}
\end{eqnarray}
This is a system of three equations on the two unknown functions
$h_0$ and $h_1$, but it is not over-determined, as we demonstrate
in what follows.
Acting on   both sides of (\ref{1.2c})
by the operator  $\sigma^2\partial_{\eta }$
and using (\ref{1.2a}) in order to eliminate  $\partial_{\eta}h_0$
in the obtained expression, one arrives eventually at
\begin{equation}
\partial_rF=4k\sigma^2\partial_rh_1-2\left( \partial^2_rh_1\right) \sigma \partial_r\sigma .
\label{1.3}
\end{equation}
Here $F$ is given
\begin{equation}
F=\sigma^2\left( \partial^2_{\eta }h_1-\partial^2_rh_1 \right)+
l\left( l+1\right) h_1  -2\sigma^2\partial_{\eta }\left( h_1
{\partial_{\eta }a\over a} \right) .
\label{1.4}
\end{equation}
Analogously, replacing  $\partial_{\eta}h_0$ by  $\partial_rh_1$
-- thanks to (\ref{1.2a}) -- in  Eq. (\ref{1.2b}),
one can obtain
\begin{equation}
F=-2\sigma \partial_r\sigma \partial_rh_1 +2h_1.
\label{1.5}
\end{equation}
Comparison of  (\ref{1.3}) and (\ref{1.5}) allows one to conclude, that
the two equations are equivalent --
modulo terms in $h_1$ which are
$r$-independent -- if  the compatibility condition
\begin{equation}
-2\partial_rh_1\left[ \partial_r\left( \sigma \partial_r \sigma \right)
-1 +2k\sigma^2 \right] =0.
\label{1.6}
\end{equation}
is satisfied. It is a straighforward exercise to check out that
the expression within the square bracket vanishes for all Friedman models.
If one assumes in addition compact support  of initial data, then the above mentioned
ambiguity disappears and one can conclude
that the two equations,  (\ref{1.3}) and (\ref{1.5}),
are equivalent. The whole dynamical content of the evolution
 is described by Eqs. (\ref{1.4}) and (\ref{1.5}),
\begin{equation}
\sigma^2\left( \partial^2_{\eta }h_1-\partial^2_rh_1 \right) +
l\left( l+1\right) h_1   -2\sigma^2\partial_{\eta }\left( h_1
{\partial_{\eta }a\over a} \right) = -2\sigma \partial_r\sigma \partial_rh_1 +2h_1.
\label{1.7}
\end{equation}
Insert now $Q= h_1/(\sigma a)$ into equation (\ref{1.7}); this yields
\begin{equation}
\partial_{\eta }^2Q-\partial_r^2Q +{l(l+1) \over \sigma^2}Q
-kQ -  {\partial_{\eta }^2a\over a}Q =0.
\label{1.8}
\end{equation}
One can show, using Friedman equations, that in the radiation epoch
$\partial_{\eta }^2a =-ka$ while in the dust epoch
$\partial_{\eta }^2a =-ka+4\pi \rho a^3/3$. Thence
the gravitational wave evolves during radiation era according to
the equation
\begin{equation}
\partial_{\eta }^2Q-\partial_r^2Q +{l(l+1) \over \sigma^2}Q=0.
\label{1.10}
\end{equation}
During the matter (and, in particular, dust) dominated epoch, in turn,
the gravitational waves satisfy
\begin{equation}
\partial_{\eta }^2Q-\partial_r^2Q +{l(l+1) \over \sigma^2}Q
-{4\pi \rho a^2\over 3}Q =0.
\label{1.11}
\end{equation}

\section{Analysis.}

During the radiation epoch the Huygens principle holds true.
Indeed, the form of the equation  (\ref{1.10})  is exactly  that of the
electromagnetic fields  in Friedman spacetimes. Therefore  its general
solution has the form (see, for instance, \cite{MWK})
\begin{equation}
\phi_{l}(r,\eta) = \sigma^l\,\underbrace{ {\partial_{r}}\,{\frac{1}
{\sigma}}\,{\partial_{r}}\,{\frac{1}{\sigma}}\cdots
{\partial_{r}}\,\bigl(\frac{f+g}{\sigma}}_{l}\bigr)
\label{2.2}
\end{equation}
where  the free functions $f$ and $g$ depend on the combinations
$r-\eta $ or $r+\eta $, respectively.
Clearly, from this form of the solution,
compact pulses of radiation  have to move within space-time
regions bounded by a pair of null cones $\eta = {^+_-}r +const$.
In particular shocks have to move along null cones.

Now we shall discuss the evolution of gravitational waves in the
matter-dominated era.
Assume   initial data of compact support $\sigma (r) \le \sigma (r_0)$
at the beginning  $\eta_0$ of this epoch.
Thus  at a time $\eta $ the   radius of the support
of $Q$ is bounded from above by $\sigma (r_0+\eta )$; assume in addition
 $\sigma (r+\eta )<\pi /2$ and $k/a^2<<H^2$
 in the case of the $k=1$ universe.
Let $R(\eta )=a(\eta )\sigma (r+\eta )$ be the related areal radius.
Define the  critical-length  parameter  $\lambda (\eta )
\equiv \sqrt{3l(l+1)/(4\pi \rho )}$.
Let us remark that, in the case  of the flat   universe,
$\lambda (\eta )$ is of the order of
the quantity $\sqrt{2l\left( l+1\right)}/H$
(where $H$ is the Hubble constant). This is related to the Hubble radius
and   for low valus of $l$  is
of the order of the radius of the visible part of the universe.
In the case of closed and open universes  $\lambda (\eta )$ can be
bounded from above or below, respectively, by $\sqrt{2l\left( l+1\right)}/H$
-- again, for small $l$, being close to
the Hubble radius $1/H$.
One can write, taking into account that  the product $\rho (\eta )
a^3(\eta )$
is constant during the matter dominated era, a sequence of inequalities
relating the ratio $N(\eta )$ of the two radii, taken at succesive moments of conformal time,
\begin{eqnarray}
&&N(\eta_1)\equiv {R^2(\eta_1 )\over \lambda^2(\eta_1 )}\le
\nonumber \\
&&{R^2(\eta_2 )\over \lambda^2(\eta_2 )}{a(\eta_2)\over a(\eta_1)}
=  N(\eta_2)\left( 1+z(1,2)\right) .
\label{2.4}
\end{eqnarray}
Here $z(1, 2)\equiv {a(\eta_2)\over a(\eta_1)}$ is the cosmological redshift
and $\eta_0 \le  \eta_1<  \eta_2 $.

The ratio of the last two terms of Eq. (\ref{1.11}) is
bounded from above by
$ 4\pi \rho a^2(\eta ) \sigma^2(r+\eta ) /\left( 3l(l+1)\right) $.
This can be expressed  as
$N(\eta )=R^2(\eta )/\lambda^2(\eta )$.
Clearly, if $N(\eta_2 ) \left( 1+z(1,2)\right) <<1$, then
$N( \eta_1 )<<1$. That suggests  that for $\eta \in \left( \eta_0 ,
\eta_2\right) $
the evolution is well described by the equation (\ref{1.10}) and the solution is
close to that given by (\ref{2.2}). Therefore there is no backscatter.
And conversely, if the above product is at least of the order of unity,
 then there can exist a backscattered part,
with  characteristics possibly allowing the identification of the type of the universe.
This heuristic reasoning can be made rigorous, as outlined below.

One can easily  check that the energy
\begin{equation}
E(Q)\equiv {1\over 2}\int_0^{\infty }dr\left[ \left( \partial_{\eta }Q\right)^2
+\left( \partial_rQ\right)^2 +{l(l+1) \over \sigma^2}Q^2\right]
\label{2.5}
\end{equation}
is conserved by the evolution equation (\ref{1.11}).
Split the sought solution of (\ref{1.11}) as   $Q=Q_0+x$, where $Q_0$ satisfies
Eq. (\ref{1.10}) and  initially $Q=Q_0, \partial_{\eta }Q=\partial_{\eta }Q_0$.
Thence $x$ solves the equation
\begin{equation}
\partial_{\eta }^2x-\partial_r^2x +  {l(l+1) \over \sigma^2}x
= {4\pi \rho a^2\over 3}\left(  x+Q_0\right)
\label{2.6}
\end{equation}
and satisfies homogeneous initial conditions  $x=\partial_{\eta }x=0$. The energy
$E(x)$ of the field $x$ satisfies the equation
\begin{equation}
{d\over d\eta }E(x)=  \int_0^{\infty }dr {4\pi \rho a^2\over 3}
\left( Q_0+x\right) \partial_{\eta }x.
\label{2.7a}
\end{equation}
One notices that $\sigma \left( r_0+\eta \right) \ge \sigma \left( r\right) $
for all points $r$  lying within  the support of $Q_0$.

Therefore  equation (\ref{2.7a}) implies the inequality
\begin{equation}
{d\over d\eta }E(x)\le {4\pi \rho a^2\over 3\sqrt{l\left( l+1\right) }}
\sigma \left( r_0+\eta \right)
\int_0^{\infty }dr  {\sqrt{l(l+1)}\over \sigma (r)}
|\left( Q_0+x\right) \partial_{\eta }x|.
\label{2.7}
\end{equation}
Now, making use of the Schwarz inequality and bounding the relevant factors on the
the right hand side of (\ref{2.7}) by energies  $E(x)$ and $E(Q_0)$, dividing both
sides by $\sqrt{E(x)}$ and rearranging the resulting inequality, one obtains
\begin{equation}
 e^{\alpha \left( \eta \right) }{d\over d\eta }\left( e^{-\alpha \left( \eta \right) }\sqrt{E(x,
 \eta )}\right) \le  {d\alpha \over d\eta } \sqrt{E(Q_0)},
\label{2.8}
\end{equation}
where
\begin{equation}
\alpha = \int_{\eta_0}^{\eta }d\tilde \eta
{4\pi \rho a^2\over 3\sqrt{l\left( l+1\right) }}
\sigma \left( r_0+\tilde \eta \right) .
\label{2.8a}
\end{equation}
In the case of flat and open universes one has  $4\pi \rho a^2/ 3\le H^2a^2/2$,
while for the closed universe there is an approximate equality
$4\pi \rho a^2/ 3\approx H^2a^2/2$, under the condition imposed earlier.
Notice now that $d\eta Ha^2=da$; change the integration variable and
observe that $\sigma \left( r_0+\eta \right) $ and $H(\eta )$
are increasing with $\eta $, in order to get the estimate
\begin{equation}
\alpha \le \sigma \left( r_0+\eta \right) \left( a\left( \eta  \right)
-  a\left( \eta_0  \right) \right){ H\left( \eta \right) \over
2\sqrt{l\left( l+1\right) }} .
\label{2.9}
\end{equation}
This can be written concisely in terms of the areal radius and the
critical parameter,
\begin{equation}
\alpha \le \beta \equiv { R\left( \eta \right) \over
\lambda \left( \eta \right) \sqrt{2}} .
\label{2.10}
\end{equation}
Finally, one arrives at the inequality
\begin{equation}
  E(x, \eta ) \le   E(Q_0) \left( e^{\beta }
  -1\right)^2 .
\label{2.11}
\end{equation}
Similar estimates can be obtained for the energy
 $E\left( \partial_rx\right) $ and of higher
derivatives of $x$. Employing these,
one can use standard Sobolev estimates \cite{Berger}
 in order to show that the field $x$ (and its derivatives, if needed)
can be pointwise small, under the condition that initial data at the time
$\eta_0=0$ have compact support and that  $R\left( \eta \right) /\left(
 \lambda \left( \eta  \right) \right) <<1$.

Let us remark that we omit boundary terms coming from $r=0$,
without  discussing  subtleties related with the behaviour of
 solutions at $r=0$, when integrating by parts. This is justified by the fact
that "good" initial data of our $1+1$ evolution problem --
i. e., those inherited from the initial $1+3$ formulation of the
evolution of gravitational waves -- take care of themselves at $r=0$ and they
give no contribution in relevant boundary terms that might appear above.

The compactness condition  $R(r, \eta )<<\lambda
\left( \eta \right)  $ means that
the areal size of the radiation pulse
is small comparing to the critical radius $\lambda \left(   \eta \right) $.
This in turn implies,
using the Fourier transforms \cite{Bracewell}, that in this (no-backscatter)
case the spectrum is dominated by high frequencies.
The compelling  explanation of the observed correlation
(dominant high frequency)--(no backscatter) is that
short length-waves do not undergo backscatter.
If the above condition is not satisfied,  then we do not have analytic
arguments that would necessitate strong backscatter, but a preliminary
numerical analysis suggests that this indeed is the case.

In conclusion, short gravitational waves - with the adjective "short" meaning
a wavelength  much smaller than the Hubble radius for the $k=0,-1$ universes
--  propagate  without backscatter.   Longer gravitational waves,
in particular those with the  wavelengths  being of the order of the radius
of the horizon (for $k=0,-1$ Friedman models, and also for
the $k=1$ model, under additional conditions)   are expected to backscatter
during the matter dominated era.
This effect can  manifest most strongly in the superhorizon scale.


\begin{thebibliography}{10}

\bibitem{Weinberg}  S. Weinberg, Gravitation and Cosmology, Wiley and Sons
1972.
\bibitem{Hadamard}    J. Hadamard {\it Lectures on Cauchy's problem
in linear partial differential equations}, Yale University Press, Yale,
New Haven 1923.
\bibitem{back} C. V. Vishveshwara, {\it Nature} {\bf 227}, 936(1970).
\bibitem{price} R. Price, {\it Phys. Rev. } {\bf D5}, 2419(1972).
\bibitem{kark} J. Karkowski, Z. \'Swierczy\'nski and E. Malec,
{\it Classical and Quantum Gravity} {\bf 21}, 1303(2004).
\bibitem{RW} T. Regge and  J. A. Wheeler, {\it Phys. Rev. } {\bf 108},
1063(1957).
\bibitem{Zerilli} F. J. Zerilli, {\it Phys. Rev.} {\bf D2}, 2141(1970).
\bibitem{MWK} E. Malec, G. Wyl\c e\.zek and J. Karkowski,
{\it General Relativity and Gravitation} {\bf 36} 2151(2004).
\bibitem{Berger} M. S. Berger, {\it Nonlinearity and functional analysis}, Academic Press,
New York, San Francisco, New York 1977.
\bibitem{Bracewell} R. Bracewell,
{\it The Fourier Transform and its applications},
McGraw-Hill Book Company 1965.

\end{thebibliography}
\end{document}